\documentclass[
  ,final            % use final for the camera ready runs
  ,numberedheadings % uncomment this option for numbered sections
  ,numbers,sort&compress
  ]
  {aipproc}
\layoutstyle{6x9}
\usepackage{write-up}
\begin{document}
%%%%%%%%%%%%%%%%%%%%%%%%%%%%%%%%%%%%%%%%%%%%
%% FRONTMATTER
%%%%%%%%%%%%%%%%%%%%%%%%%%%%%%%%%%%%%%%%%%%%
\title{Single-Spin Asymmetries and Transversity%
%%%  \thanks{%
%%%    The XV International Symposium on High-Energy Spin Physics
%%%    (Brookhaven, 9--14 Aug.\ 2002)
%%%  }%
}
%------------------------------------------------------------------------------%
\author{Philip G. Ratcliffe}{
  address={Dip.to di Scienze CC.FF.MM.,
           Univ. degli Studi dell'Insubria---sede di Como, Italy\\
           and INFN---sezione di Milano, Italy},
  email={philip.ratcliffe@uninsubria.it},
  homepage={http://www.unico.it/~pgr},
  thanks={The \emph{Insubri} were a Celtic tribe, originally from across the
          Alps, who in the $5^\mathrm{th.}$ century B.C. settled roughly the
          area known today as Lombardy.},
}
%------------------------------------------------------------------------------%
%------------------------------------------------------------------------------%
\begin{abstract}
  A pedagogical introduction to \acr{SSAs} and transversity is presented.
  Discussion in some detail is made of certain aspects of \acr{SSAs} in
  lepton--nucleon and in hadron--hadron scattering and the role of p\acs{QCD}
  and evolution in the context of transversity.
\end{abstract}

\maketitle
%------------------------------------------------------------------------------%
\section{Preamble}

%%%\subsection{Single-Spin Asymmetries}

Single-spin asymmetries are one of the oldest forms of high-energy spin
measurement, the reason being accessibility: the only requirement is
\emph{either} beam \emph{or} target polarised, for $\Lambda^0$ production
neither is necessary! However, after early interest (due to large experimental
effects), a theoretical ``dark age'' descended: p\acs{QCD} had apparently nothing
to say, save that such asymmetries are zero! We now know that the rich
phenomenology is matched by a richness of the theoretical framework: the main
topic of my talk.

One might argue the inapplicability of p\acs{QCD} to existing \acr{SSA} data
owing to the low $Q^2$ accessed while there are several non-p\acs{QCD} models
that can explain some (though not all) of the data. Examples may be found in
\cite{Andersson:1979wj, DeGrand:1981pe, Barni:1992qn, Soffer:1992am}. However,
I shall examine \acr{SSAs} purely from within the p\acs{QCD} framework.

%%%\subsection{Transversity}

Transversity too has a long history: the concept (though not the term) was
introduced in \citeyear{Ralston:1979ys} by \citeauthor*{Ralston:1979ys} via the
Drell--Yan process. The \acr{LO} anomalous dimensions were first calculated by
\citet{Baldracchini:1981uq} but forgotten. They were recalculated by
\citet*{Artru:1990zv} and it turns out that they had also been obtained by
various groups as part of the $g_2$ evolution \cite{Kodaira:1979ib,
Antoniadis:1981dg, Bukhvostov:1985rn, Ratcliffe:1986mp}. A complete
classification of chirally-odd densities including transversity, is due to
\citet*{Jaffe:1992ra}. However, as yet there are no experimental data on
transversity. This is owing to the inaccessibility (discussed later) of
transversity in inclusive \acr{DIS}.

After introducing \acl{SSAs} and transversity, I shall discuss \acr{SSAs} in
lepton-nucleon and hadron-hadron scattering in some detail and close with a few
brief comments and conclusions. A large part of what follows is taken from the
\emph{Physics Reports} by \citet*{Barone:2001sp} and from a forthcoming book by
\citet*{Barone:2002b2}. Thus, much credit is due to my two collaborators.

%------------------------------------------------------------------------------%
\section{Introduction}

%%%\subsection{Single-Spin Asymmetries}

Generically, \acr{SSAs} reflect correlations of the form
$\Vec{s}\cdot(\Vec{p}\vprod\Vec{k})$, where $\Vec{s}$ is a polarisation vector
while $\Vec{p}$ and $\Vec{k}$ are particle/jet momenta. An example is $\Vec{s}$
a (transverse) target
\newpage
\noindent polarisation, $\Vec{p}$ a beam direction, and $\Vec{k}$ that of a
final-state particle. Thus, polarisations in \acr{SSAs} will typically be
transverse (but see later). Transforming the basis from transverse spin to the
more familiar helicity,
$\ket{\uparrow/\downarrow}=\textfrac1{\sqrtno2\,}[\ket{+}\pm\I\ket{-}]$, such
an asymmetry takes on the (schematic) form
\begin{equation}
  \mathcal{A}
  \binsep\sim
  \frac{\braket{\uparrow}{\uparrow}-\braket{\downarrow}{\downarrow}}
       {\braket{\uparrow}{\uparrow}+\braket{\downarrow}{\downarrow}}
  \binsep\sim
  \frac{2\Im\braket{+}{-}}{\braket{+}{+}+\braket{-}{-}} \; .
\end{equation}
The presence of both $\ket+$ and $\ket-$ in the numerator implies a spin-flip
amplitude while the precise form indicates interference between spin-flip and
non-flip amplitudes, with a non-trivial relative phase difference.

It was soon realised \cite{Kane:1978nd} that in the Born approximation and
massless limit a gauge theory, such as \acr{QCD}, cannot furnish either
requirement since fermion helicity is conserved and tree diagrams are real.
Quoting from \cite{Kane:1978nd}: ``\dots\ \emph{observation of significant
polarizations in the above reactions would contradict either \acr{QCD} or its
applicability.}'' \, Later, however, examining the three-parton correlators
related to $g_2$, \citet*{Efremov:1985ip} found a way out: the mass scale
relevant to spin flip is not that of a current quark but the hadron and the
two-loop nature of the diagrams can give rise to an imaginary part.
Nonetheless, it was a while before the complexity of the new structures was
fully exploited (see, \eg, \cite{Qiu:1991pp, Qiu:1992wg}).

%%%\subsection{Transversity}

\begin{figure}[hbtp]
  \centering
  \includegraphics[width=40mm,bb=158 571 310 681,clip]{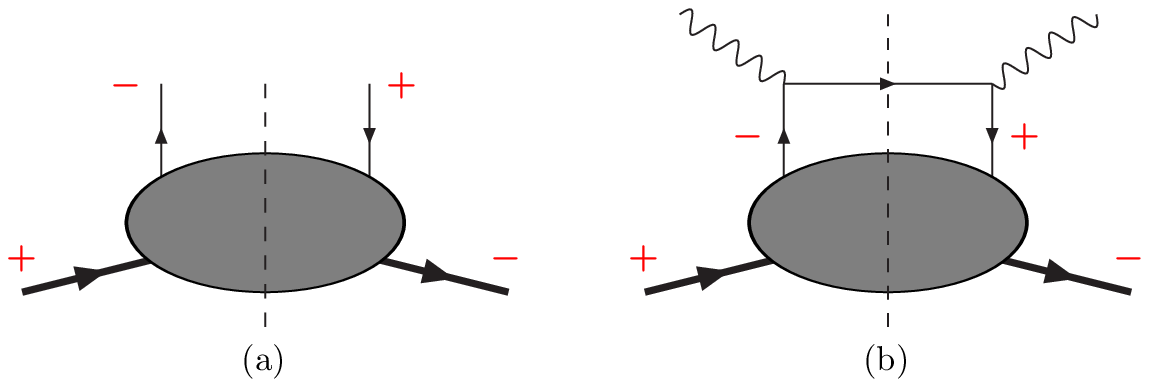}
  \hspace{6em}
  \includegraphics[width=40mm,bb=336 571 486 679,clip]{figures/chirality}
  \caption{(a) Chirally-odd hadron--quark amplitude,
           (b) chirality-flip forbidden \acs{DIS} diagram.
  }
  \label{fig:chirality-flip}
\end{figure}

Transversity is the third twist-two partonic density. At this point it is
important to make the distinction between partonic \emph{densities}---$q(x)$,
$\DL{q}(x)$, $\DT{q}(x)$, \dots\ and \acs{DIS} \emph{structure
functions}---$F_1$, $F_2$, $g_1$, $g_2$, \dots\ \, In the leading-twist
unpolarised and helicity-dependent cases there is a simple connection between
the two: \acr{DIS} structure functions are weighted sums of partonic densities;
in contrast, there is no \acr{DIS} transversity structure function and $g_2$
does not correspond to a partonic density.
%%%
%%%\subsection{Chirality Flip}
%%%
The absence of transversity in \acr{DIS} is illustrated in
Fig.~\ref{fig:chirality-flip}.
Note that chirality flip is not a problem if the quarks connect to different
hadrons, as in \acl{DY} processes.

%%%\section{\acs{QCD} Evolution of Transversity}

%%%\subsection{Twist Basics and Operators}

The three twist-two structures are then:
\begin{subequations}
\begin{eqnarray}
  f(x) &=&
  \int \! \frac{\d\xi^-}{4\pi} \e^{\I xP^+ \xi^-}
  \langle PS |
    \anti\psi(0)
      \gamma^+
    \psi(0, \xi^-, \Vec{0}_\perp)
  | PS \rangle \, ,
\\
  \DL{f}(x) &=&
  \int \! \frac{\d\xi^-}{4\pi} \e^{\I xP^+ \xi^-}
  \langle PS |
    \anti\psi(0)
      \gamma^+ \gamma_5
    \psi(0, \xi^-, \Vec{0}_\perp)
  | PS \rangle \, ,
\\
  \DT{f}(x) &=&
  \int \! \frac{\d\xi^-}{4\pi} \e^{\I xP^+ \xi^-}
  \langle PS |
    \anti\psi(0)
      \gamma^+ \gamma^1 \gamma_5
    \psi(0, \xi^-, \Vec{0}_\perp)
  | PS \rangle \, .
\end{eqnarray}
\end{subequations}
The $\gamma_5$ matrix signals spin dependence while the extra $\gamma^1$ matrix
in $\DT{f}(x)$ signals the chirality-flip that precludes transversity
contributions in \acr{DIS}.

%%%\subsection{Ladder Diagram Summation}

\begin{figure}[hbtp]
  \centering
  \includegraphics[width=0.35\textwidth,bb=20 60 290 320,clip]
                  {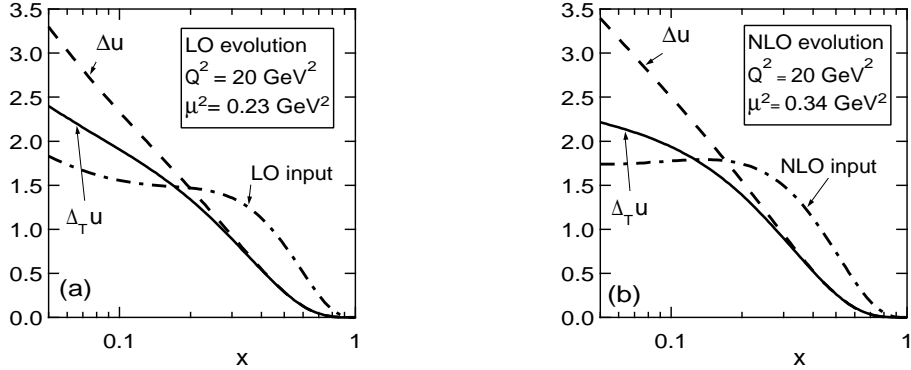}
  \hspace{4em}
  \includegraphics[width=0.35\textwidth,bb=290 60 560 320,clip]
                  {figures/hkk-fig6}
  \caption{The $Q^2$-evolution of $\DT{u}(x,Q^2)$ and $\DL{u}(x,Q^2)$ compared at
    (a) \acs{LO} and (b) \acs{NLO}; from \cite{Hayashigaki:1997dn}.}
  \label{fig:evolution}
\end{figure}

For somewhat similar reasons the \acr{LO} \acr{QCD} evolution of transversity
is non-singlet like: quark--gluon mixing would require a chirality-flip in a
\begin{figure}[hbtp]
  \centering
  \begin{fmffile}{gluonrung1}
    \begin{fmfgraph*}(30,30)
      \fmfpen{thick}
      \fmfbottom{i1,o2}
      \fmftop{o1,i2}
      \fmf{fermion,label=\boldmath{$\red+$},l.side=left}{i1,v1}
      \fmf{fermion,label=\boldmath{$\red+$},l.side=left}{v1,o1}
      \fmf{fermion,label=\boldmath{$\red-$},l.side=left}{i2,v2}
      \fmf{fermion,label=\boldmath{$\red-$},l.side=left}{v2,o2}
      \fmffreeze
      \fmf{gluon}{v1,v2}
    \end{fmfgraph*}
  \end{fmffile}
  \hspace{4em}
  \begin{fmffile}{fermionrung1}
    \begin{fmfgraph*}(30,30)
      \fmfpen{thick}
      \fmfbottom{i1,o2}
      \fmftop{o1,i2}
      \fmf{fermion}{i1,v1}
      \fmf{gluon}{o1,v1}
      \fmf{gluon}{v2,i2}
      \fmf{fermion}{v2,o2}
      \fmffreeze
      \fmf{fermion,label=\boldmath{$\red+$},l.side=left}{i1,v1}
      \fmf{fermion,label=\boldmath{$\red-$},l.side=left}{v2,o2}
      \fmf{fermion,label={\red?}}{v1,v2}
    \end{fmfgraph*}
  \end{fmffile}
  \caption{Left, transversity evolution kernel;
           right, disallowed gluon--fermion mixing.}
  \label{fig:kernels}
\end{figure}
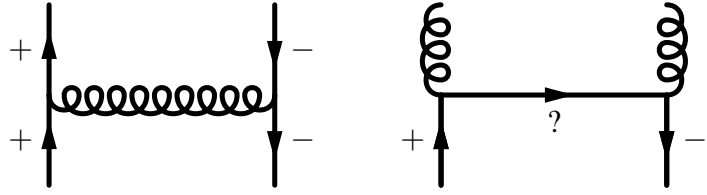
%%%
%%%\subsection{\acs{LO} and \acs{NLO} Evolution}
%%%
quark propagator---see Fig.~\ref{fig:kernels}. The \acr{LO} quark--quark
splitting functions are then:
\begin{subequations}
\begin{eqnarray}
  P_{qq}^{(0)}
  &=&
  \CF \left( \frac{1+x^2}{1-x\,} \right)_+ \, ,
\\
  \DL{P}_{qq}^{(0)}
  &=&
  P_{qq}^{(0)}
  \hspace*{1em}
  \text{(by helicity conservation),}
\\
  \DT{P}_{qq}^{(0)}
  &=&
  \CF \left[ \left( \frac{1+x^2}{1-x\,} \right)_+ - 1 + x \right]
  \; = \;
  P_{qq}^{(0)}(x) - \CF (1-x) \, .
\end{eqnarray}
\end{subequations}
Note that the first moments of both $P_{qq}^{(0)}$ and $\DL{P}_{qq}^{(0)}$
vanish (leading to conservation laws and sum rules), but not so of
$\DT{P}_{qq}^{(0)}$. The effects of evolution are shown in
Fig.~\ref{fig:evolution}.

%%%\section{The Soffer Bound}

\begin{wrapfigure}{r}{0.3\textwidth}
  \centering
  \vspace*{-1.5ex}
  \includegraphics[width=0.3\textwidth]{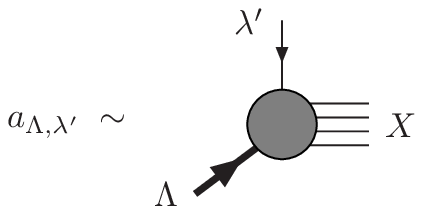}
\end{wrapfigure}
By considering hadron--parton helicity amplitudes (see the figure alongside),
\citet{Soffer:1995ww} constructed an interesting bound involving transversity.
Taking into account all relevant symmetries there are only two independent
amplitudes, in terms of which all three densities are expressed:
\begin{subequations}
\begin{eqnarray}
    f(x)
    &\propto&
    \Im (\mathcal{A}_{++,++} + \mathcal{A}_{+-,+-})
    \binsep\propto
    \textstyle
    \sum_X (a_{++}^* a_{++} + a_{+-}^* a_{+-}) \, ,
  \\[0.5ex]
    \DL{f}(x)
    &\propto&
    \Im (\mathcal{A}_{++,++} - \mathcal{A}_{+-,+-})
    \binsep\propto
    \textstyle
    \sum_X ( a_{++}^* a_{++}- a_{+-}^* a_{+-}) \, ,
  \\[0.5ex]
    \DT{f}(x)
    &\propto&
    \Im \mathcal{A}_{+-,-+}
    \hphantom{(\null - \mathcal{A}_{+-,+-})}
    \binsep\propto
    \textstyle
    \sum_X a_{--}^* a_{++} \, .
\end{eqnarray}
\end{subequations}
A straight-forward Schwartz-type inequality: $\sum_X|a_{++}\pm{}a_{--}|^2\ge0$
then translates into $f_+(x)\ge|\DT{f}(x)|$ or $f(x)+\DL{f}(x)\ge2|\DT{f}(x)|$,
which is precisely the Soffer bound. Notice that it involves all three
leading-twist structures.

%------------------------------------------------------------------------------%
\section{A \acs{DIS} Definition for Transversity}

The other twist-two densities are naturally defined via \acr{DIS}, where the
parton picture is formulated and many model calculations performed. When
translated to \acr{DY} processes, large $K$ factors appear
$\sim\text{O}(\pi\alpha_s)$. At RHIC energies this corresponds
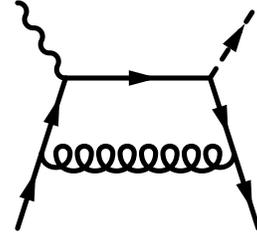
\begin{wrapfigure}{r}{40mm}
  \centering
  \vspace*{-0.5ex}
  \begin{fmffile}{higgs1}
    \begin{fmfgraph*}(40,30)
      \fmfpen{thick}
      \fmfleft{i1,i2}
      \fmfright{o1,o2}
      \fmf{fermion,tension=1.0}{i1,u1,v1}
      \fmf{fermion,tension=0.5}{v1,v2}
      \fmf{fermion,tension=1.0}{v2,u2,o1}
      \fmf{photon}{i2,v1}
      \fmf{scalar}{v2,o2}
      \fmffreeze
      \fmf{gluon}{u1,u2}
      \fmfv{label=\rnode{NA}{}}{v2}
    \end{fmfgraph*}
  \end{fmffile}
  \hfill
  \caption{Hypothetical Higgs--photon interference mechanism.}
  \label{fig:Higgs-photon}
\end{wrapfigure}
to a $\sim30\%$ correction, at EMC/SMC nearly $100\%$. Pure \acr{DY}
coefficient functions are known, but are scheme dependent. Moreover, a
$\ln^2x/(1-x)$ term appears that is not found for spin-averaged or
helicity-dependent \acr{DY}. Together with recent problems arising in
connection with vector--scalar current products, this suggests an interesting
check.

One might invoke a Higgs--photon interference mechanism, which, though
experimentally hardly viable, does provide a \acs{DIS}-type definition for
transversity since the presence of a scalar vertex forces chirality flip. Care
must be taken over the extra renormalisation contribution from the scalar
vertex, which factorises into the running mass (or Higgs coupling). One then
needs to calculate diagrams of the form of Fig.~\ref{fig:Higgs-photon} (and
correspondents for \acs{DY}) in order to obtain the relevant higher-order
Wilson coefficients.

\begin{wrapfigure}{l}{0.55\textwidth}
  \centering
  \psfrag{A_DY}{$A_\mathrm{DY}$}
  \psfrag{tau}{$\tau$}
  \includegraphics[width=0.55\textwidth]{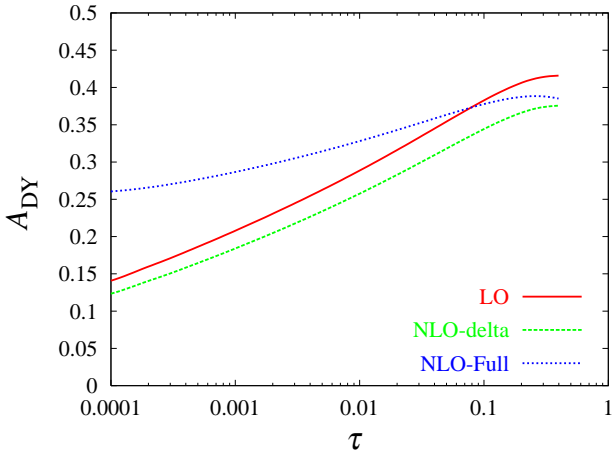}
  \caption{Transversity asymmetry (valence only) for \acs{DY};
    $\tau=Q^2/s$, $s=4{\cdot}10^4\GeV^2$.
  }
  \label{fig:Hp-evol}
\end{wrapfigure}
Fig.~\ref{fig:Hp-evol} illustrates the effect of the \acr{NLO} Wilson
coefficient \cite{Ratcliffe:0000ip}. In contrast to the helicity asymmetry
\cite{Ratcliffe:1983yj}, where the difference between the \acr{LO} and
\acr{NLO} is small (the large coefficient of the $\delta$-function is identical
in the numerator and denominator), here there are important differences between
the spin-averaged denominator and the transversity-dependent numerator. The
principal culprits are the $\delta$-function coefficient and the new
$\frac{\ln^2x}{1-x}$ term. The \acs{DIS}--\acs{DY} ``transformation'' coefficients
for the unpolarised and transversity cases are:
\begin{subequations}
\begin{eqnarray}
  C^f_{q,DY}-2C^f_{q,\acr{DIS}} &=& \frac{\alpha_s}{2\pi}\frac43
  \left[
    \frac3{(1-z)_+} + 2(1+z^2)\left(\frac{\ln(1-z)}{1-z}\right)_+
    - 6 - 4z
  \right.
  \nonumber
\\
  && \hspace*{50mm}
  \left.\vphantom{\left(\frac{\ln()}{1}\right)_+}\null
    + \left(\frac43\pi^2+1\right) \delta(1-z)
  \right] ,
\\
  C^h_{q,DY}-2C^h_{q,DIS} &=& \frac{\alpha_s}{2\pi}\frac43
  \left[
    \frac{3z}{(1-z)_+} + 4z\left(\frac{\ln(1-z)}{1-z}\right)_+
    - 6z\frac{\ln^2z}{1-z}
    + 4(1-z)
  \right.
  \nonumber
\\
  && \hspace*{50mm}
  \left.\vphantom{\left(\frac{\ln()}{1}\right)_+}\null
    + \left(\frac43\pi^2-1\right) \delta(1-z)
  \right] .
\end{eqnarray}
\end{subequations}

%------------------------------------------------------------------------------%
\section{T-Odd Structures}

We now wish to generalise the $\Vec{k}_\perp$-integrated density functions to
include all possible correlations between the quark and parent-hadron spins,
later on we shall find we also need $\Vec{k}_\perp$-dependent generalisations.
Thus, some extra notation will be required (see \cite{Barone:2001sp}). In this
way objects like $\Delta_L^T{f}$ have a simple interpretation:
\begin{itemize}
\item
\emph{subscripts} $0,L$ and $T$ in density and fragmentation functions denote
the \emph{quark} polarisation state,
\item
\emph{superscripts} $0,L$ and $T$ denote the parent or off-spring \emph{hadron}
polarisation state.
\end{itemize}
The superscript is dropped when equal to the subscript.

\begin{wrapfigure}{l}{0.3\textwidth}
  \centering
  \vspace*{-1.5ex}
  \includegraphics[width=0.28\textwidth]{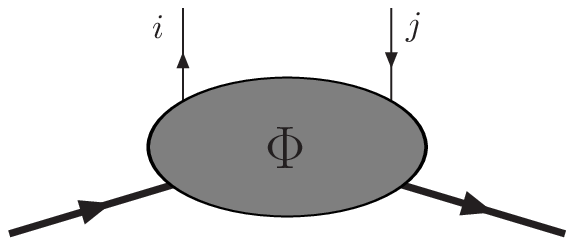}
  \caption{Quark--quark correlation matrix.}
  \label{fig:qqcormat}
\end{wrapfigure}
The aim then is to parametrise the quark--quark correlation matrix (see
Fig.~\ref{fig:qqcormat}) in the most general manner, while respecting the
natural properties of hermiticity, parity, and time-reversal invariance,
though, as we shall see later, this last may be relaxed. The most general
decomposition of $\Phi$ over a complete basis of Dirac matrices is
\begin{equation}
  \Phi (k, P, S)
  \binsep=
  \half
  \left\{
    \mathcal{S} \, \unitop +
    \mathcal{V}_\mu \, \gamma^\mu +
    \mathcal{A}_\mu \gamma_5 \gamma^\mu +
    \I \mathcal{P}_5 \gamma_5 +
    \I \, \mathcal{T}_{\mu\nu}  \, \sigma^{\mu\nu} \gamma_5
  \right\} ,
  \label{df2}
\end{equation}
where the quantities $\mathcal{S}$, $\mathcal{V}^\mu$, $\mathcal{A}^\mu$,
$\mathcal{P}_5$ and $\mathcal{T}^{\mu\nu}$ are to be constructed from the
vectors $k^\mu$, $P^\mu$ and the pseudovector $S^\mu$.

%%%\subsection{T-Odd Structures}

Relaxing $T$ invariance allows two new twist-two structures:
\begin{equation}
  \mathcal{V}^\mu
  \binsep=
  \dots + \frac1{M} \, A_1' \,
  \varepsilon^{\mu\nu\rho\sigma} \, P_\nu k_{\perp\rho} S_{\perp\sigma}
  \quad \text{and} \quad
  \mathcal{T}^{\mu\nu}
  \binsep=
  \dots + \frac1{M} \, A_2' \,
  \varepsilon^{\mu\nu\rho\sigma} \, P_\rho k_{\perp\sigma} \, .
\end{equation}
These give rise to two $\Vec{k}_\perp$-dependent $T$-odd density functions,
$f_{1T}^\perp$ and $h_1^\perp$ \citep*{Boer:1998nt}:
\begin{equation}
  \Phi^{[\gamma^+]}
  \binsep= \dots -
  \frac{\varepsilon_\perp^{ij} k_{\perp i} S_{\perp j}}{M} \,
  f_{1T}^\perp(x,\Vec{k}_\perp^2)
  \quad \text{and} \quad
  \Phi^{[ \I \sigma^{i+} \gamma_5]}
  \binsep= \dots -
  \frac{\varepsilon_\perp^{ij} k_{\perp j}}{M} \,
  h_{1}^\perp(x,\Vec{k}_\perp^2) \, .
\end{equation}

The partonic interpretation is as follows. The density $f_{1T}^\perp$ relates
to the number density of unpolarised quarks in a transversely polarised
nucleon:
\begin{eqnarray}
  \mathcal{P}_{q/N \uparrow}   (x, \Vec{k}_\perp) -
  \mathcal{P}_{q/N \downarrow} (x, \Vec{k}_\perp)
  &=&
  \mathcal{P}_{q/N \uparrow} (x,   \Vec{k}_\perp) -
  \mathcal{P}_{q/N \uparrow} (x, - \Vec{k}_\perp)
  \nonumber
\\
  &=&
  -2 \frac{| \Vec{k}_\perp |}{M} \, \sin(\phi_k - \phi_S)\,
  f_{1T}^\perp(x, \Vec{k}_\perp^2) \, .
\end{eqnarray}
The $T$-odd density $h_1^\perp$ measures quark transverse polarisation in an
unpolarised hadron:
\begin{equation}
  \mathcal{P}_{q \uparrow/N}   (x, \Vec{k}_\perp) -
  \mathcal{P}_{q \downarrow/N} (x, \Vec{k}_\perp)
  \binsep=
  - \frac{| \Vec{k}_\perp |}{M} \, \sin(\phi_k - \phi_s)
  \, h_1^\perp(x, \Vec{k}_\perp^2) \, .
\end{equation}
It is convenient to define two quantities $\Delta_0^T{f}$ and $\DT^0{f}$
(related to $f_{1T}^\perp$ and $h_1^\perp$ respectively) by absorbing the
factors $|\Vec{k}_\perp|/M$:
\begin{equation}
  \Delta_0^T f(x, \Vec{k}_\perp^2)
  \binsep\equiv
  -2 \frac{| \Vec{k}_\perp |}{M} \, f_{1T}^\perp(x, \Vec{k}_\perp^2)
  \quad \text{and} \quad
  \DT^0 f(x, \Vec{k}_\perp^2)
  \binsep\equiv
  - \, \frac{| \Vec{k}_\perp |}{M} \, h_{1}^\perp(x, \Vec{k}_\perp^2) \, .
\end{equation}

The question now arises as to why we should entertain such $T$-odd quantities
at all.
%%%
%%%\subsection{T-Odd Justification}
%%%
There are various attitudes: \citet{Anselmino:1998yz} (among others) advocate
initial-state effects, which prevent implementation of na{\"\i}ve time-reversal
invariance. The suggestion is that the colliding hadrons interact strongly with
non-trivial relative phases, akin to those arising from final-state effects. An
alternative has been proposed by \citet{Anselmino:2002yx}: they apply a general
argument on time reversal for particle multiplets suggested by
\citet{Weinberg:1995mt}. If the internal structure of hadrons is described at
some low momentum scale by a chiral lagrangian, time reversal might be realised
in a ``non-standard'' manner that could mix the multiplet components. According
to this approach, the $u$ ($d$) density transforms into the $d$ ($u$) density,
and time-reversal invariance simply establishes a relation between the $u$ and
$d$ sectors.

Finally, \citet{Collins:2002kn} has recently reconsidered his proof of the
vanishing of $f_{1T}^\perp$ and $h_1^\perp$, based on the field-theoretical
expressions of the two densities. He noticed that on reinstating the link
operators into quark--quark bi-locals the densities do not simply change sign
under $T$; a future-pointing Wilson line becomes past-pointing. Consequently,
time-reversal invariance, does not constrain $f_{1T}^\perp$ and $h_1^\perp$ to
be zero, but relates processes probing Wilson lines in opposite directions.
\citeauthor{Collins:2002kn} thus predicts the \citeauthor{Sivers:1990cc}
asymmetry, \eg, to have opposite signs in \acr{DIS} and in \acr{DY}.

%------------------------------------------------------------------------------%
\section{Lepton-Nucleon Scattering}

%%%Topics:
%%%\begin{itemize}
%%%  \item Exclusive Processes % Skewed densities
%%%  \item Single Longitudinal-Spin Asymmetries
%%%\end{itemize}

%%%\subsection{Exclusive Processes}

\begin{wrapfigure}{r}{0.3\textwidth}
  \vspace*{-1ex}
  \includegraphics[width=0.26\textwidth,bb=334 572 451 670,clip]{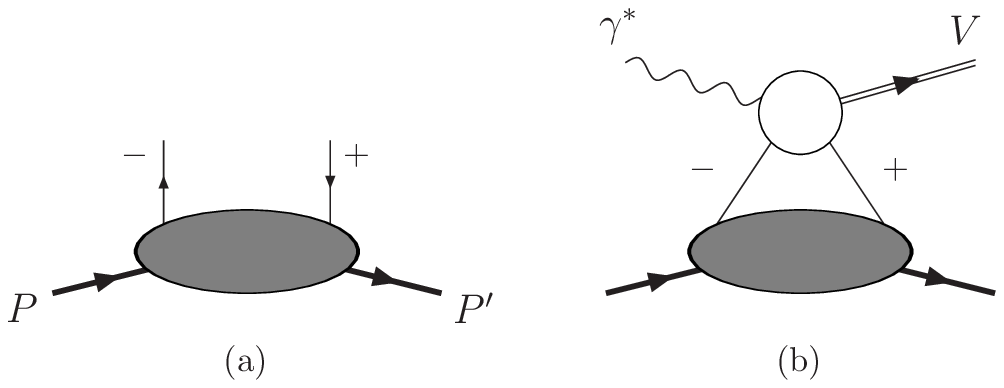}
  \caption{Exclusive production of vector-mesons.}
  \label{fig:skewed}
\end{wrapfigure}
One might hope to access transversity through exclusive leptoproduction of
vector mesons (see Fig.~\ref{fig:skewed} alongside). However,
\citet{Mankiewicz:1998uy} showed that the chirally-odd contribution to
vector-meson production is actually zero at \acr{LO} in $\alpha_s$.
\citet{Diehl:1998pd} and \citet{Collins:1999un} later extended this, observing
that the chirally-odd contribution vanishes due to angular-momentum and
chirality conservation in the hard scattering and so holds at leading twist to
all orders in $\alpha_s$. Thus, exclusive vector-meson leptoproduction cannot
probe (off-diagonal) transversity densities.

%%%\subsection{Single Longitudinal-Spin}

The cross-section for production off a longitudinally polarised target is
\cite{Kotzinian:1997wt}:
\begin{eqnarray}
  \frac{\d^5\sigma(\lambda_N)}{\d{x} \d{y} \d{z} \d^2\Vec{P}_{h\perp}}
  &=& -
  \frac{4\pi\alpha_\text{em}^2s}{Q^4} \,
  \lambda_N \sum_a e_a^2 \,
  x (1-y) \, \sin(2 \phi_h)
  \nonumber
\\
  && \hspace{1em} \null \times
  I\left[
    \frac{2
          (\hat{\Vec{h}}{\cdot}\Vec\kappa_\perp)
          (\hat{\Vec{h}}{\cdot}\Vec{k}_\perp) -
          \Vec\kappa_\perp{\cdot}\Vec{k}_\perp}
         {M M_h} \,
    h_{1La}^\perp(x,\Vec{k}_\perp) \,
    H_{1a}^\perp(z,\Vec\kappa_\perp)
  \right] .
\end{eqnarray}
Transversity is not present here, but the asymmetry does depend on the Collins
function $H_1^\perp\propto\sin(2\phi_h)$, also on a $\Vec{k}_\perp\!$-dependent
density function $h_{1L}^\perp$.

%%%\subsection{Leptoproduction Summary}

Summarising, in the context of semi-inclusive \acr{DIS} there are four
candidate leading-twist reactions to determine $\DT{f}$: namely, inclusive
leptoproduction of
\begin{enumerate}
\item
a transversely polarised hadron from a transversely polarised target;
\item
an unpolarised hadron from a transversely polarised target;
\item
two hadrons from a transversely polarised target;
\item
a spin-one polarised or unpolarised hadron from a transversely polarised
target.
\end{enumerate}

%------------------------------------------------------------------------------%
\section{Hadron-Hadron Scattering}

%%%Topics:
%%%\begin{itemize}
%%% \item Single-Particle Production
%%% \begin{itemize}
%%%  \item Transverse-Momentum Effects
%%%  \item Twist-Three Effects
%%% \end{itemize}
%%% \item Drell--Yan
%%%\end{itemize}

%%%\subsection{Single-Hadron Production}

\begin{wrapfigure}{r}{0.55\textwidth}
  \centering
  \includegraphics[width=0.55\textwidth]{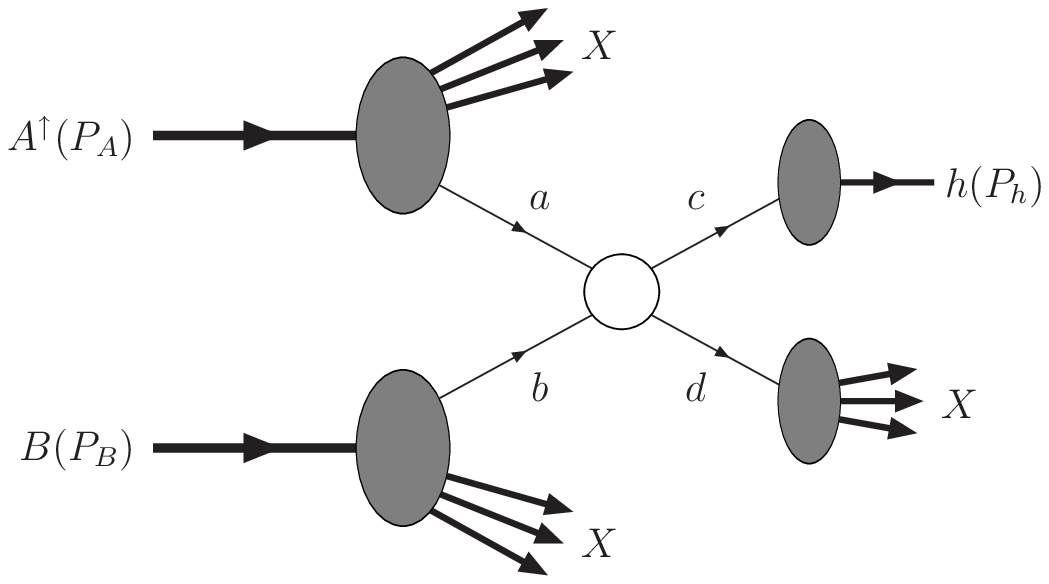}
  \caption{Single-hadron production with a transversely polarised target.}
  \label{fig:single}
\end{wrapfigure}
Let us now examine single-hadron production with a transversely polarised
target. The process is exemplified in Fig.~\ref{fig:single} alongside: $A$ is
transversely polarised and the unpolarised (or spinless) hadron $h$ is produced
at large transverse momentum $\Vec{P}_{hT}$, therefore p\acs{QCD} is
applicable. In typical experiments $A$ and $B$ are protons while $h$ is a pion.
According to the factorisation theorem, the differential cross-section for the
reaction may be written formally as
\begin{equation}
  \d\sigma
  \binsep=
  \sum_{abc} \sum_{\alpha\alpha'\gamma\gamma'} \,
  \rho^a_{\alpha'\alpha} \;
  f_a(x_a) \otimes
  f_b(x_b) \otimes
  \d\hat\sigma_{\alpha\alpha'\gamma\gamma'} \otimes
  \mathcal{D}_{h/c}^{\gamma'\gamma}(z) \, ,
\end{equation}
$f_a$ ($f_b$) is the density of parton $a$ ($b$) inside hadron $A$ ($B$),
$\rho^a_{\alpha\alpha'}$ is the parton $a$ spin density matrix,
$\mathcal{D}_{h/c}^{\gamma\gamma'}$ is the fragmentation matrix of parton $c$
into hadron $h$ and $\d\hat\sigma/\d\hat{t}$ is the elementary cross-section:
\begin{equation}
  \left(
    \frac{\d\hat\sigma}{\d\hat{t}}
  \right)_{\alpha\alpha'\gamma\gamma'}
  \binsep=
  \frac1{16\pi\hat{s}^2} \, \frac12 \, \sum_{\beta\delta}
  \mathcal{M}_{\alpha\beta\gamma\delta} \,
  \mathcal{M}^*_{\alpha'\beta\gamma'\delta} \, .
\end{equation}

\begin{wrapfigure}{r}{0.3\textwidth}
  \centering
  \vspace*{-1.5ex}
  \includegraphics[width=0.25\textwidth]{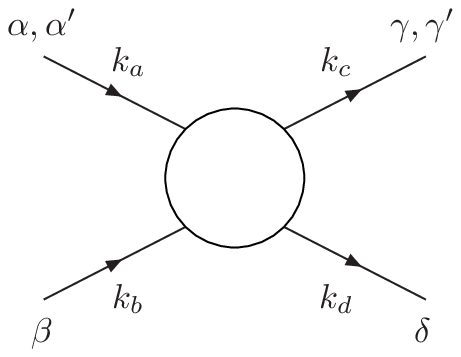}
  \caption{Partonic hard scattering amplitude.}
  \label{fig:partonamp}
\end{wrapfigure}
\noindent Here $\mathcal{M}_{\alpha\beta\gamma\delta}$ is the amplitude for the
hard partonic process, displayed in Fig.~\ref{fig:partonamp}. For an
unpolarised produced hadron, the off-diagonal elements of
$\mathcal{D}_{h/c}^{\gamma\gamma'}$ vanish, \ie,
$\mathcal{D}_{h/c}^{\gamma\gamma'}\propto\delta_{\gamma\gamma'}$. Helicity
conservation then implies $\alpha=\alpha'$ and thus there can be no dependence
on the spin of hadron $A$. Consequently, all \acr{SSAs} must vanish.

%%%\subsection{Transverse Motion and \acs{SSAs}}

To avoid this conclusion, intrinsic quark transverse motion or higher-twist
effects must be invoked; this can be done in three different ways:
\begin{enumerate}
\item
$\Vec\kappa_T$ in hadron $h$ allows $\mathcal{D}_{h/c}^{\gamma\gamma'}$ to be
non-diagonal (a fragmentation $T$-odd effect), the \citeauthor{Collins:1993kk}
effect \cite{Collins:1993kk};
\item
$\Vec{k}_T$ in hadron $A$ implies that $f_a(x_a)$ should be replaced by the
$\mathcal{P}_a(x_a,\Vec{k}_T)$, which may depend on the spin of hadron $A$ (a
density $T$-odd effect), the \citeauthor{Sivers:1990cc} effect
\cite{Sivers:1990cc};
\item
$\Vec{k}'_T$ in hadron $B$ implies that $f_b(x_b)$ should be replaced by
$\mathcal{P}_b(x_b,\Vec{k}'_T)$; a transverse spin of parton $b$ in the
unpolarised hadron $B$ may then couple to the transverse spin of parton $a$ in
$A$ (a density $T$-odd effect), see \cite{Boer:1999mm}.
\end{enumerate}
It should be stressed that all these intrinsic-$\Vec\kappa_T$, -$\Vec{k}_T$, or
-$\Vec{k}'_T$ effects are $T$-odd. Note too that when intrinsic quark
transverse motion is taken into account, the \acr{QCD} factorisation theorem is
not proven.

Assuming, for discussion purposes, factorisation to be valid, the cross-section
is
\begin{eqnarray}
  E_h \, \frac{\d^3\sigma}{\d^3\Vec{P}_h} &=&
  \sum_{abc} \, \sum_{\alpha\alpha'\beta\beta'\gamma\gamma'} \,
  \int \!\!\! \d{x}_a \int \!\!\! \d{x}_b
  \int \!\!\! \d^2\Vec{k}_T \int \!\!\! \d^2\Vec{k}'_T
  \int \!\!\! \d^2\Vec\kappa_T \,
  \frac1{\pi z}
  \nonumber
\\
  && \hspace{0em} \null \times
  \mathcal{P}_a(x_a, \Vec{k}_T) \, \rho^a_{\alpha'\alpha} \,
  \mathcal{P}_b(x_b, \Vec{k}'_T) \, \rho^b_{\beta'\beta}
  \left(
    \frac{\d\hat\sigma}{\d\hat{t}}
  \right)_{\alpha\alpha'\beta\beta'\gamma\gamma'}
  \mathcal{D}_{h/c}^{\gamma'\gamma}(z, \Vec\kappa_T) \, ,
\end{eqnarray}
where
\begin{equation}
  \left(
    \frac{\d\hat\sigma}{\d\hat{t}}
  \right)_{\alpha\alpha'\beta\beta'\gamma\gamma'}
  \binsep=
  \frac1{16\pi\hat{s}^2} \, \sum_\delta
  \mathcal{M}_{\alpha\beta\gamma\delta}
  \mathcal{M}^*_{\alpha'\beta'\gamma'\delta} \, .
\end{equation}
The \citeauthor{Collins:1993kk} mechanism requires intrinsic quark transverse
motion inside the produced hadron $h$ while neglecting all other quark
transverse momenta (the spin of $A$ is along $y$):
\begin{eqnarray}
  E_h \, \frac{\d^3\sigma( \Vec{S}_T)}{\d^3\Vec{P}_h} -
  E_h \, \frac{\d^3\sigma(-\Vec{S}_T)}{\d^3\Vec{P}_h}
  &=& -
  2 \, | \Vec{S}_T | \, \sum_{abc}
  \int \! \d{x}_a \int \! \frac{\d{x}_b}{\pi z} \int \! \d^2\Vec\kappa_T
  \qquad
  \nonumber
\\
  && \hspace{-5em} \null \times
  \DT{f}_a(x_a) \, f_b(x_b) \,
  \Delta_{TT} \hat\sigma(x_a, x_b, \Vec\kappa_T) \,
  \DT^0 D_{h/c}(z, \Vec\kappa_T^2) \, ,
\end{eqnarray}
where $\Delta_{TT}\hat\sigma$ is a partonic spin-transfer asymmetry. The
\citeauthor{Sivers:1990cc} effect relies on $T$-odd density functions and
predicts a form
\begin{eqnarray}
  E_h \, \frac{\d^3\sigma( \Vec{S}_T)}{\d^3\Vec{P}_h} -
  E_h \, \frac{\d^3\sigma(-\Vec{S}_T)}{\d^3\Vec{P}_h}
  &=&
  | \Vec{S}_T | \, \sum_{abc} \int \! \d{x}_a \int \! \frac{\d{x}_b}{\pi z}
  \int \! \d^2\Vec{k}_T
  \nonumber
\\
  && \hspace{-3em} \null \times
  \Delta_0^T f_a(x_a, \Vec{k}_T^2) \, f_b(x_b) \,
  \frac{\d\hat\sigma(x_a, x_b, \Vec{k}_T)}{\d\hat{t}} \, D_{h/c}(z) \, ,
\end{eqnarray}
where $\Delta_0^T{f}$ (related to $f_{1T}^\perp$) is a $T$-odd density.
Finally, the effect studied by \citeauthor{Boer:1999mm} gives rise to an
asymmetry involving the other $T$-odd density $\DT^0f$ (related to
$h_1^\perp$):
\begin{eqnarray}
  E_h \, \frac{\d^3\sigma( \Vec{S}_T)}{\d^3\Vec{P}_h} -
  E_h \, \frac{\d^3\sigma(-\Vec{S}_T)}{\d^3\Vec{P}_h}
  &=& -
  2 | \Vec{S}_T | \, \sum_{abc}
  \int \! \d{x}_a \int \! \frac{\d{x}_b}{\pi z} \int \! \d^2\Vec{k}'_T
  \nonumber
\\
  && \hspace{-5em} \null \times
  \DT{f}_a(x_a) \, \DT^0 f_b(x_b, \Vec{k'}_T^2) \,
  \Delta_{TT} \hat\sigma'(x_a, x_b, \Vec{k}'_T) \, D_{h/c}(z) \, ,
\end{eqnarray}
where $\Delta_{TT}\hat\sigma'$ is a partonic initial-spin correlation.

As already mentioned, \citet{Efremov:1982sh} pointed out that \acr{SSAs} can
arise in p\acs{QCD} at higher twist via gluonic poles in diagrams involving
$qqg$ correlators. Such asymmetries were evaluated by \citeauthor*{Qiu:1991pp},
who studied direct photon production \cite{Qiu:1991pp, Qiu:1992wg} and hadron
production \cite{Qiu:1998ia}. The extension to chirally-odd contributions was
made by \citet{Kanazawa:2000hz, Kanazawa:2000kp}. The results may be summarised
in
\begin{eqnarray}
  \d\sigma &=&
  \sum_{abc}
  \left\{ \vphantom{D_{h/c}^{(3)}}
    G_F^a(x_a, y_a) \otimes f_b(x_b) \otimes
    \d\hat\sigma \otimes D_{h/c}(z)
  \right.
  \nonumber
\\[-1ex]
  && \hspace{3em} \null +
  \DT{f}_a(x_a) \otimes E_F^b(x_b, y_b) \otimes
  \d\hat\sigma' \otimes D_{h/c}(z)
  \nonumber
\\
  && \hspace{5em} \null +
  \left.
    \DT{f}_a(x_a) \otimes f_b(x_b) \otimes
    \d\hat\sigma'' \otimes D_{h/c}^{(3)}(z)
  \right\} .
\end{eqnarray}
The first term (not containing transversity) is a chirally-even mechanism
studied by \citeauthor*{Qiu:1998ia}, the second term is the chirally-odd
contribution analysed by \citeauthor*{Kanazawa:2000hz}, and the third contains
a twist-three fragmentation function $D_{h/c}^{(3)}$.

%%%\subsection{Drell--Yan at Twist Three}

Admitting twist-three contributions, the \acr{SSA} in \acr{DY} is
\cite{Boer:1998bw}
\begin{eqnarray}
  A_{T}^\text{DY} &=&
  | \Vec{S}_{1\perp} | \,
  \frac{2\sin2\theta}{1+\cos^2\theta} \,
  \sin(\phi-\phi_{S_1}) \,
  \frac{M}{Q}
  \nonumber
\\
  && \hspace{4em} \null \times
  \frac{\sum_a e_a^2
         \left[
           x_1 \, \widetilde{f}_T^a(x_1)\, \bar{f}_a(x_2) +
           x_2 \, \DT{f}_a(x_1) \, \bar{h}_a(x_2)
         \right]
       }
       {\sum_a e_a^2 \, f_a(x_1) \bar{f}_a(x_2)} \; ,
  \label{dytwist34}
\end{eqnarray}
\begin{wrapfigure}{r}{0.35\textwidth}
  \centering
  \vspace*{-2ex}
  \includegraphics[width=0.32\textwidth,bb=208 542 403 690,clip]{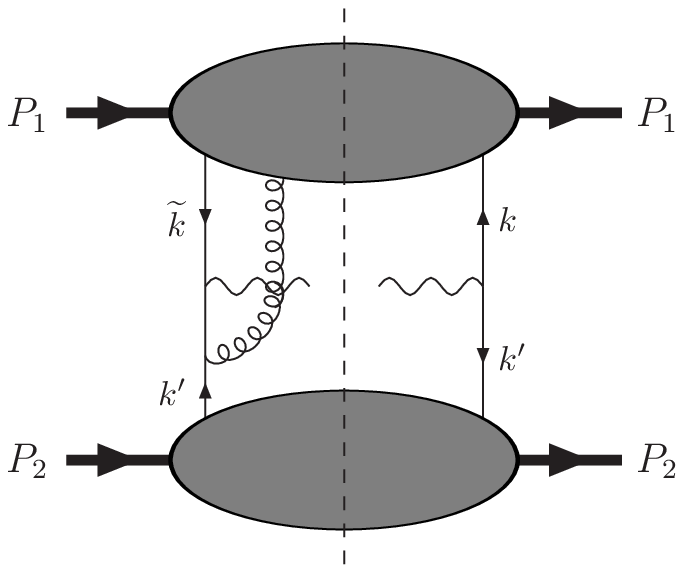}
  \caption{A twist-three gluon-pole contribution to \acs{DY}.}
  \label{fig:DY-T3}
\end{wrapfigure}
where $\widetilde{f}_T(x)$ and $\bar{h}(x)$ are twist-three $T$-odd density
functions. The existence of such $T$-odd density functions has been advocated
by \citet{Boer:1999mm} to explain an anomalously large $\cos2\phi$ term seen in
unpolarised \acr{DY} data. As presented, such contributions would require
initial-state interactions---this may be considered unlikely.
\citet{Hammon:1997pw} have shown that \acr{SSAs} may arise from gluonic poles
in twist-three multiparton correlation functions (see Fig.~\ref{fig:DY-T3}
alongside). The corresponding \acr{SSA} is then
\begin{eqnarray}
  A_{T}^\text{DY}
  &\propto&
  | \Vec{S}_{1\perp} | \,
  \frac{2\sin2\theta}{1+\cos^2\theta} \,
  \sin(\phi - \phi_{S_1}) \, \frac{M}{Q}
  \nonumber
\\
  && \hspace{4em} \null \times
  \frac{\sum_a e_a^2 \,
        [
          G_F^a(x_1, x_1) \, \bar{f}_a(x_2) +
          \DT{f}_a(x_1) \, E_F^a(x_2, x_2)
        ]
       }
       {\sum_a e_a^2 \, f_a(x_1) \bar{f}_a(x_2)} \; .
\end{eqnarray}
Comparing this with the previous expression we may identify
\begin{subequations}
\begin{eqnarray}
    f_T^\text{eff}(x)
    &\sim&
    \textstyle
    G_F(x,x)
    \binsep\sim
    \int \! \d{y} \, \Im G_A^\text{eff}(x,y) \, ,
  \\[0.5ex]
    h^\text{eff}(x)
    &\sim&
    \textstyle
    E_F(x,x)
    \binsep\sim
    \int \! \d{y} \, \Im E_A^\text{eff}(x,y) \, .
\end{eqnarray}
\end{subequations}
Thus, $T$-odd functions at twist three, can explain $A_T^\text{DY}$ via
quark--gluon interactions, without initial-state effects.

%------------------------------------------------------------------------------%
\section{Conclusions}

The study of single-spin asymmetries has become a very complex area of
high-energy spin physics. A plethora of new structure and fragmentation
functions has opened the way to explaining much existing phenomenology.
However, in order to distinguish and separate out the various mechanisms
proposed, a large amount of diverse high-energy data will be necessary and it
is difficult (if not indeed irrelevant and even misleading) to single out at a
few key experiments. In other words, all new data will be very welcome.

%------------------------------------------------------------------------------%
%%%\bibliographystyle{pigroaye}
%%%\bibliography{pigrostr,pigropgr,pigrodbf,pigroxrf}
%%%\end{document}
%------------------------------------------------------------------------------%

%------------------------------------------------------------------------------%
%%%\newpage
%%%\tableofcontents
%------------------------------------------------------------------------------%
\end{document}